\documentstyle[12pt,psfig]{article}

\newcommand{\NP}[1]{ Nucl.\ Phys.\ {#1}}

\newcommand{\PR}[1]{Phys.\ Rev.\ { #1}}
\newcommand{\PRL}[1]{ Phys.\ Rev.\ Lett.\ { #1}}

\newcommand{\diag}{\mbox{diag}}

\newcommand{\be}{\begin{equation}}
\newcommand{\ee}{\end{equation}}
\newcommand{\ba}{\begin{eqnarray}}
\newcommand{\ea}{\end{eqnarray}}

\newcommand{\da}{^\dagger}

\newcommand{\f}{\frac}

\newcommand{\dd}{\displaystyle}

\newcommand{\lr}[1]{{ \left( \, #1 \, \right) }}
\newcommand{\gs}{g'_5 \, \! ^2}

\begin{document}
\baselineskip=20pt

\begin{center}
  {\Large  \bf The Equivalence Theorem for gauge boson scattering
in a five
dimensional Standard Model}

\vspace{.5cm} { Stefania DE CURTIS$^a$,
Daniele DOMINICI$^{a}$, Jos\'e R. PELAEZ$^{a,b}$
}

\emph{ $^a$ INFN,
Sezione di Firenze and
Dip. di Fisica, Univ. degli Studi, Firenze,
Italy.}

\emph{ $^b$Dept. de F\'{\i}sica Te\'orica II,
  Univ. Complutense, 28040 Madrid, Spain.}
\end{center}

\vspace{-.5cm}
\noindent
\rule{\textwidth}{.1mm}
\begin{abstract}
\noindent
We present an Equivalence Theorem for the longitudinal
components of the gauge bosons in a
 compactified five dimensional extension of the Standard Model,
whose spontaneous symmetry breaking is driven either by  one
Higgs in the bulk or by one on a brane
or by both together. We also
show some implications for the unitarity bounds on Higgs masses.
\end{abstract}
\vspace{-.5cm}
\rule{\textwidth}{.1mm}

\section{Introduction}

There has been recently a growing interest in theories with large
and extra large dimensions motivated by the multidimensional
unification of gravitational,
 strong and electroweak interactions through string theory.
Special attention has been devoted to the brane picture where
ordinary matter lives in four dimensions while gravity propagates
in the bulk. Specific models relating the solution  to the
hierarchy problem to the existence of a large volume for the $n$
extra dimensions \cite{Arkani-Hamed:1998rs} or to an exponential
warp factor in a five dimensional (5D) non-factorizable metric
\cite{rs}  have been suggested; as in all Kaluza Klein (KK)
theories  the compactification process produce a tower of
graviton and scalar excitations, whose phenomenology has been studied in
\cite{Giudice:1998ck}.

In addition  there are realizations where also the gauge
interactions feel some extra  dimensions, parallel to the brane:
supersymmetric  5D Standard Model (SM) extensions 
have been proposed, where the supersymmetry breaking scale is related to the
compactification scale which therefore turns out to be in the TeV
range \cite{Antoniadis:1990ew}. Many formal and phenomenological aspects of these models
have been investigated; in particular these models contain KK
towers of excitations of the $W$, $Z$ gauge bosons, of the photon
and possibly of the Higgs. Lower bounds from the electroweak
precision data on the compactification scale of these models, when
fermions are localized on the brane or in different points of the
bulk, are in the range of 2-5 TeV
\cite{Masip:1999mk,Muck:2001yv}. These bounds become much weaker,
300-400 GeV, when
all particles live in the bulk
\cite{Barbieri:2000vh}.

Recently the unitarity of
5D Yang Mills theories has been considered
\cite{SekharChivukula:2001hz}, proving a
theorem similar to the standard Equivalence Theorem (ET) \cite{ET,Lee:1977eg}
that relates at high energies the longitudinal components of gauge bosons to 
their associated Goldstone bosons (GB). In the unbroken
extra dimensional Yang Mills case, what has been shown is the
equivalence of 
longitudinal KK gauge bosons $V^\mu_{(n)}$ and their
corresponding $V^5_{(n)}$ components of the 5D
gauge fields.

The aim of this work is to 
show this equivalence in the  case
of spontaneously broken 5D extensions of the SM. 
The main subtlety in the proof is that, the usual
SM would-be GB can mix with 
KK states.
In section 2, using the
formalism of the non supersymmetric 5D SM
and its compactification to four dimensions,
we identify those GB and  $V^5_{(n)}$ KK combinations 
$\hat G^V_{(n)}$ that couple
to the gauge field derivatives in the gauge fixing term.
From here proceeds the standard proof of the ET
 in the $R_\xi$ gauge that relates the
scattering amplitudes between longitudinal KK gauge bosons and 
those of their corresponding $\hat G^V_{(n)}$: $
  T(\hat V^{\mu }_{L\,(m)},\hat V^{\mu }_{L\,(n)},\ldots)
\simeq  T(\hat G^V_{(m)},\hat G^V_{(n)},\ldots) +O(M_k/E) $, $M_k$
being the biggest one of the masses of the KK gauge bosons.
Finally, we illustrate the use of the ET 
to calculate the 
$W^{+}_{L\,(m)}W^{-}_{L\,(n)}\rightarrow W^{+}_{L\,(p)}W^{-}_{L\,(q)}$ 
scattering amplitudes for the 
 channel. These amplitudes are relevant
 to investigate how the partial wave unitarity limit on
 the mass of the Higgs is modified  by extra dimensions.
 Some aspects of this problem have been
addressed in particular by studying the effect of the radion in
the $HH$ \cite{Bae:2000pk} or in the
$W_L^+W_L^-$ amplitudes \cite{Han:2001xs}. 
In section 3 we calculate the amplitudes in the case with just  one
 Higgs  in the bulk, in section 4, with just one Higgs on the brane and
finally, in section 5 we study the general case with
one Higgs on the brane and another one
 in the bulk.

\section{The Equivalence Theorem for 5D fields}
\label{section2}
 We consider a minimal 5D extension
of the SM with two scalar fields, compactified on the segment
$S^1/Z_2$, of length $\pi R$, in which the $SU(2)_L$ and $U(1)_Y$
gauge fields and the Higgs field $\Phi_1$   propagate in the bulk
while  the Higgs field $\Phi_2$ lives on the brane at $y=0$. The
Lagrangian of the gauge Higgs sector is given by (see
\cite{Muck:2001yv}
 for a review)
\ba \int_{0}^{2 \pi R} dy\int
dx\,{\cal L}(x,y) &=&\int_{0}^{2 \pi R} dy\int dx\,\Big\{ - \frac
1 4 B_{MN}B^{MN} - \frac 1 4 F^a_{MN}F^{aMN}+{\cal L}_{GF}(x,y) 
\nonumber\\
&+&
(D_M \Phi_1)^\dagger (D^M \Phi_1)+\delta(y)
(D_\mu \Phi_2)^\dagger (D^\mu \Phi_2)- V(\Phi_1,\Phi_2)
\Big\},
\label{kinlagrangian}
\ea
where $M=\mu,5$, $B_{MN}$, $F_{MN}^a$ are
the $U(1)_Y$ and $SU(2)_L$ field strengths and
 $a$ is the $SU(2)$ index. Note that $\Phi_1$ has
energy dimension 3/2, whereas $\Phi_2$ has dimension 1.
The covariant derivative is defined as
$D_M=\partial_M - i g_5 A^a_M\tau^a/2 - i g_5'B_M/2$.
For simplicity we will consider a Higgs potential symmetric under
the discrete symmetry $\Phi_2\rightarrow -\Phi_2$, which is given by
\begin{eqnarray}
\label{scalarpotential}
V(\Phi_1,\Phi_2) \!\!\!&=&\!\! \mu_1^2 \, ( \Phi_1\da \Phi_1 ) \, + \,
\lambda^{(5)}_1 \, ( \Phi_1\da \Phi_1 )^2 \,
                            + \, \delta(y) \, \Big[\,
\frac{1}{2}\,\mu_2^2 \, ( \Phi_2\da \Phi_2 ) \,
+ \,
\frac{1}{2}\,\lambda_2 \, ( \Phi_2\da \Phi_2 )^2 \nonumber\\
&+&\!\!\!
\frac{1}{2}\,\lambda^{(5)}_3 \, ( \Phi_1\da \Phi_1 ) ( \Phi_2\da \Phi_2 ) \,
   + \, \frac{1}{2}\,\lambda^{(5)}_4
\, ( \Phi_1\da \Phi_2 ) ( \Phi_2\da \Phi_1 )\,
   + \, \lambda^{(5)}_5\, ( \Phi_1\da \Phi_2 )^2\,
+ \, {\rm h.c.}\, \Big], 
\end{eqnarray}
where the dimensionalities
of these couplings are: 1 for $\mu_1$ and $\mu_2$,
 -1 for
$\lambda_1^{(5)},\lambda^{(5)}_3,\lambda^{(5)}_4$ and
$\lambda^{(5)}_5$,
whereas $\lambda_2$ is dimensionless. Also for simplicity, we will
require $\lambda^{(5)}_3+\lambda^{(5)}_4+2\lambda^{(5)}_5=0$,
which ensures that the minimum of the potential corresponds to the
constant configuration $\Phi_1=(0,v_1/\sqrt{4\pi R})$,
$\Phi_2=(0,v_2/\sqrt{2})$, where $v_1^2\equiv -2 \pi R
\mu_1^2/\lambda_1^{(5)}$ and $v_2^2\equiv -\mu_2^2/\lambda_2$. In
this way, the Higgs fields are expanded in the standard form 
\be
\Phi_1(x,y)=\left (
\begin{array}{c}
 \frac{i}{\sqrt{2}}(\omega^1-i\omega^2)\\
\dd { \frac{1}{\sqrt{2}}(\f {v_1}{\sqrt{2\pi R}}+ h_1-i \omega^3) }
\\
\end{array}
\right), \,\,\,
\Phi_2(x)=\left (
\begin{array}{c}
  \frac{i}{\sqrt{2}}(\pi^1-i\pi^2)\\
\dd { \frac{1}{\sqrt{2}}( {v_2}+ h_2-i \pi^3) }
\\
\end{array}
\right),
\ee
where following the
standard two Higgs notations $v_1=v \cos\beta$,
$v_2=v\sin\beta$ are the vacuum expectation values of the scalar fields
and $v^2=(\sqrt{2} G_F)^{-1}$.
For brevity we will use the notation $c_\beta\equiv \cos\beta$,
$s_\beta\equiv \sin\beta$.

The gauge fixing Lagrangian ${\cal L}_{GF}(x,y)$ is
\begin{eqnarray}
&&{\cal L}_{GF}(x,y)=- \f {1}{2 \xi}(F^a(A^a))^2 - \f {1}{2 \xi}(F(B))^2
\label{GF5},\\
&&  F^a(A^a)=\partial_\mu A^{a\,\mu }-\xi\left[\partial_5 A^a_5-
\f{g_5v c_\beta}{2\sqrt{2\pi R}}\omega^a-\f{g_5v s_\beta}{2}\pi^a\delta(y)
\right],
\nonumber\\
&&  F(B)=\partial_\mu B^{\mu }-\xi\left[\partial_5 B_5+
\f{g'_5v c_\beta}{2\sqrt{2\pi R}}\omega^3+\f{g'_5v s_\beta}{2}\pi^3\delta(y)
\right],\nonumber
\end{eqnarray}
where,
in order to avoid a gauge dependent mixing angle between
the physical $Z$ and the photon, we have chosen the same $\xi$ parameter
for the $A^{a\,\mu}$ and $B^\mu$ fields. 
Let us now recall that the fields living in the bulk have a
Fourier expansion, which is:
\begin{eqnarray}
  X(x,y)=\f{1}{\sqrt{2\pi R}}X_{(0)}(x)+\f{1}{\sqrt{\pi R}}
\sum_{n=1}^{\infty}\cos\left(\f{n y}{R}\right)X_{(n)}(x),
\end{eqnarray}
for $X=A_\mu^a, B_\mu, \omega^a, h_1$,  whereas for $Y=A_5^a, B_5$ it is
\begin{equation}
    Y(x,y)=\f{1}{\sqrt{\pi R}}
\sum_{n=1}^{\infty}\sin\left(\f{n y}{R}\right)Y_{(n)}(x).
\end{equation}
Note that  the condition
$\lambda^{(5)}_3+\lambda^{(5)}_4+2\lambda^{(5)}_5=0$
yields a diagonal Higgs mass matrix: $m^2_{h_1(0)}=2v_1^2 \lambda_1$,
$m^2_{h_2}=2v_2^2 \lambda_2$, $m^2_{h_1(n)}=2v_1^2 \lambda_1+n^2/R^2$,
 where $\lambda_1=\lambda_1^{(5)}/(2\pi R)$.

Similarly to the SM case in four dimensions, we define
the following charged and neutral field combinations
$W^{\pm}_M \, =
\lr{ A^1_M \, \mp \, i \, A^2_M } /\sqrt{2} $,
$Z_M \,  = \lr{ g_5
\, A^3_M \, - \, g'_5 B_M }/\sqrt{g_5^2 + \gs}$, 
$A_M \,  = \lr{ g'_5 \, A^3_M \, + \, g_5 B_M }/\sqrt{g_5^2
+ \gs}$.
After integrating out the compactified fifth dimension $y$,
the mass matrix ${\cal M}_V^2$
of the gauge bosons and their KK excitations
has the following $(N+1)\times(N+1)$ generic form
(with $N\rightarrow\infty$):
\begin{equation}
 (V_{(0)},V_{(1)},V_{(2)},\ldots)\left(
\begin{array}{cccc}
m^2 +d_0^2& \sqrt{2} m^2 &  \sqrt{2} m^2 &\ldots\\
 \sqrt{2} m^2 &2m^2+ d_1^2 &   \sqrt{2} m^2 &\ldots\\
 \sqrt{2} m^2 &  \sqrt{2} m^2 & 2m^2+ d_2^2   &\ldots\\
\vdots & \vdots & \vdots & \ddots
\end{array}
\right)
\left(\begin{array}{c}
V_{(0)}\\
V_{(1)}\\
V_{(2)}\\
\vdots
\end{array}
\right),
\label{genericM}
\end{equation}
where $m^2=m_V^2 s_\beta^2$ and
$d_0=m_V c_\beta$, $d_n=\sqrt{(n/R)^2+d_0^2}$. In particular,
for $V=W^\pm_\mu$, $m_W=gv/2$,  whereas for $V=Z_\mu$,
$m_Z=\sqrt{g^2+g'^2}v/2$,
with $g=g_5/\sqrt{2\pi R}$ and $g'=g'_5/\sqrt{2\pi R}$.
Note that for the photon $m=m_A=0$, the mass matrix
is already diagonal, the photon has zero mass
and for its associated KK states the masses
are given by $m_{A(n)}=n/R$.

For the $V=W_\mu^\pm, Z_\mu$ case, ${\cal M}_V^2$ is diagonal when 
 $s_\beta=0$ and when $s_\beta\neq0$
it  has the following eigenvalue equation
\begin{eqnarray}
  \sqrt{m_{V(n)}^2-d_0^2}&=&
\frac{m^2}{ \sqrt{m_{V(n)}^2- d_0^2}}
\left(1+2\sum_{i=1}^N\frac{m_{V(n)}^2-d_0^2}
{m_{V(n)}^2-d_i^2} \right)
\nonumber\\
&&
\stackrel{N\rightarrow\infty}{\longrightarrow}\pi\,m^2\,R
\cot\left( \pi\,R  \sqrt{m_{V(n)}^2-d_0^2} \right),
\label{eigenvalueeq}
\end{eqnarray}
so that it can be diagonalized with 
$P_V^t{\cal M}_V^2 P_V=\diag\{m_{V(0)}^2,m_{V(1)}^2,\cdots\}$,
where 
\begin{equation}
  P_V=\left(
\begin{array}{cccc}
\dd{\frac{u_0}{\sqrt{2} }} & \ldots& \dd{\frac{u_n}{\sqrt{2}} }
&\ldots\\
\dd{ \frac{u_0(d_0^2-m_{V(0)}^2)}{d_1^2-m_{V(0)}^2} } & \ldots&
\dd{ \frac{u_n(d_0^2-m_{V(n)}^2)}{d_1^2-m_{V(n)}^2} }
&\ldots\\
\vdots&\ddots&\vdots&\vdots\\
\dd{ \frac{u_0(d_0^2-m_{V(0)}^2)}{d_n^2-m_{V(0)}^2} } &\ldots&
\dd{ \frac{u_n(d_0^2-m_{V(n)}^2)}{d_n^2-m_{V(n)}^2} }
&\ldots\\
\vdots & \vdots & \vdots & \ddots
\end{array}
\right),
\end{equation}
is a $(N+1)\times(N+1)$ matrix
and
\be u_j=
\sqrt{\f 1 2 +\sum_{i=1}^{N}\left [
\frac{d_0^2-m_{V(j)}^2}{d_i^2-m_{V(j)}^2}\right ]^2 }
=\f
{2m^2}{\sqrt{m_{V(j)}^2- d_0^2}\sqrt{ m^2(1+\pi^2 R^2 m^2)+
m_{V(j)}^2 -d_0^2}},
\label{ujdef}
\ee
where we have used, in
the $N\rightarrow\infty$ limit, the following series
\be \sum_{i=1}^{\infty}
\frac {(d_0^2-m_{V(j)}^2)^2}{(d_i^2-m_{V(j)}^2)^2}=\f 1 4 \left [
-2 +\f {m_{V(j)}^2- d_0^2}{m^2}+ \pi^2 R^2 (m_{V(j)}^2- d_0^2)+\f
{(m_{V(j)}^2- d_0^2)^2}{m^4}\right ].
\ee
Thus, the gauge boson mass eigenstates are
$\hat V_{(n)}=(P_V)_{nm}V_{(m)}$.

After integrating out the fifth dimension, and by separating
the charged  and neutral field combinations, the gauge fixing conditions 
in eq.~(\ref{GF5}) become
\begin{eqnarray*}
&&{\cal L}_{GF}(x)=   -\f{1}{2\xi}\sum_{n=0}^{\infty}
\left\{ 
 2\left\vert \partial_\mu W^{+\,\mu }_{(n)}-\xi\left(
\sqrt{\f{n^2}{R^2}+m_W^2 c_\beta^2}\,\,G^+_{(n)}
-\sqrt{2}^{1-\delta_{n,0}}m_W s_\beta\,\pi^+\right)\right\vert^2
\right.
\\
&& \left.
+\left[ \partial_\mu Z^{\mu }_{(n)}-\xi\left(
\sqrt{\f{n^2}{R^2}+m_Z^2 c_\beta^2}\,\,G^Z_{(n)}
-\sqrt{2}^{1-\delta_{n,0}}m_Z s_\beta\,\pi^3\right)
\right]^2
+\left[ \partial_\mu A^{\mu }_{(n)}-\xi
{\f{n}{R}}\,\,A_{(n)}^5\right]^2
\right\},
\nonumber
\end{eqnarray*}
where
$\pi^{\pm} = \frac{1}{\sqrt{2}} \lr{ \pi^1 \, \mp \, i \, \pi^2 }$,
$\omega^{\pm} = \frac{1}{\sqrt{2}}\lr{ \omega^1 \, \mp \, i \, \omega^2 } $.
We have also defined
\begin{eqnarray}
&&G^\pm_{(0)}=-\omega^\pm_{(0)},\quad
G^\pm_{(n)}=
c^{_W}_{n} \, W^\pm_{5\,(n)}+ s^{_W}_n \,\omega^\pm_{(n)},
\; n\ge 1,\nonumber\\
&&G^Z_{(0)}=-\omega^3_{(0)},\quad
G_{(n)}^Z=
c^{_Z}_n \,\, Z_{5\,(n)}\,+ s^{_Z}_n \,\omega^3_{(n)}
, \,\; n\ge 1,
\label{Gs}
\end{eqnarray}
where $s^{_V}_n=-m_V c_\beta/ \sqrt{ n^2/R^2+m_V^2 c_\beta^2}$ and
 $c^{_V}_n=(n/R)/ \sqrt{ n^2/R^2+m_V^2 c_\beta^2}$. 
In general, for the calculations of amplitudes we would also
need the  orthogonal  combinations
\begin{eqnarray}
a^\pm_{(n)}=
-s^{_W}_{n} \, W^\pm_{5\,(n)}+ c^{_W}_n \,\omega^\pm_{(n)},\qquad
a_{(n)}^Z=
-s^{_Z}_n \,\, Z_{5\,(n)}\,+ c^{_Z}_n \,\omega^3_{(n)}
, \qquad n\ge 1.
\label{as}
\end{eqnarray}
Note that, as commented in the introduction, the usual 
GB and their KK excitations are mixed with the
KK states of $W^\pm_5$ and $Z_5$, in the gauge fixing term.

Once we have written the gauge fixing fields in the charged-neutral
basis, in order to find the GB mass
 eigenstates
it is very convenient to rewrite the gauge fixing in
a more compact matrix form including all the KK excitations.
For the sake of brevity,
we gather the gauge bosons in an N+1 dimensional vector
 ${\bf V^{\mu}}=(V^\mu_{(0)},V^\mu_{(1)},\ldots)^t$,
where now $V= W^\pm, Z, A$,
whereas the GB and the pseudoscalars  are gathered in the
N+2 dimensional vectors
${\bf G^\pm}=(\pi^\pm,G^\pm_{(0)},G^\pm_{(1)},G^\pm_{(2)},\ldots)^t$,
${\bf G^Z}=(\pi^3,G^Z_{(0)},G^Z_{(1)},G^Z_{(2)},\ldots)^t$,
${\bf G^A}=(0,0,A_{5\,(1)},A_{5\,(2)},\ldots)^t$.
With these definitions
\begin{eqnarray*}
{\cal L}_{GF}(x)=-\frac{1}{2\xi}\left\{2\left\vert
\partial_\mu {\bf W^{+ \mu}}-\xi M^W_\xi {\bf G^+}
\right\vert^2
-\left\vert
\partial_\mu {\bf Z^\mu}-\xi M^Z_\xi {\bf G^Z}
\right\vert^2
-\left\vert \partial_\mu {\bf A^{\mu }}-\xi
M^A_\xi\,\,{\bf G^A}\right\vert^2 \right\},
\end{eqnarray*}
the $(N+1)\times(N+2)$ dimensional $M^V_\xi$ matrix being
generically of the  form
\begin{equation}
  M_{V\,\xi}=\left(
\begin{array}{ccccc}
-m & d_0 &  0&  0 &\ldots\\
 -\sqrt{2} m &0& d_1 & 0 &\ldots\\
 -\sqrt{2} m &0  & 0&d_2 & \ldots\\
\vdots & \vdots & \vdots &  \vdots &\ddots
\end{array}
\right)
\end{equation}
The gauge-fixing term provides a gauge-dependent mass
term for the would-be GB
$ {\cal L}_{\xi\,mass}=-\xi {\bf G^+} {\cal M}_{W\xi}^2 {\bf G^-}
-\xi {\bf G^{Z}}^t {\cal M}_{Z\xi}^2 {\bf G^Z}/2
-\xi {\bf G^{A}}^t {\cal M}_{A\xi}^2 {\bf G^A}/2$, with
\begin{equation}
{\cal M}_{V\,\xi}^2\equiv M_{V\,\xi}^t M_{V\,\xi}=
\left(
\begin{array}{cccc}
m^2(1+\sum_{i=1}^N 2) & -m d_0 & -\sqrt{2} m d_1 &\ldots\\
-m d_0 & d_0^2 &    0 &\ldots\\
 -\sqrt{2} m d_1&0 & d_1^2 & \ldots\\
\vdots & \vdots & \vdots &  \ddots
\end{array}
\right)
\end{equation}
being a $(N+2)\times(N+2)$ matrix, whose
eigenvalues are the same as those of
${\cal M}_V^2$, plus a zero.
For the photon that matrix is already diagonal, whereas
for the $V=W_\mu^\pm,Z_\mu$ cases, it
can be diagonalized to
$Q_V^t  {\cal M}_{V\,\xi}^2 Q_V=\diag\{0,m_{V(0)}^2,m_{V(1)}^2,\cdots\}$
using the $(N+2)\times(N+2)$ orthogonal matrix:
\begin{equation}
  Q_V=\left(
\begin{array}{cccc}
\dd{q_{-1}/m}& \dd{q_{0}/m}
& \dd{q_{1}/m}&   \ldots\\
\dd{\f{q_{-1}}{d_0}}& \dd{\f{d_0q_{0}}{d_0^2-m_{V\,(0)}^2}}  &
\dd{\f{d_0q_{1}}{d_0^2-m_{V\,(1)}^2}}
&  \ldots\\
\dd{\f{\sqrt{2}q_{-1}}{d_1}}
& \dd{\f{\sqrt{2}d_1q_{0}}{d_1^2-m_{V\,(0)}^2}}
&  \dd{\f{\sqrt{2}d_1q_{1}}{d_1^2-m_{V\,(1)}^2}}
 &\ldots\\
\vdots & \vdots & \vdots &\ddots
\end{array}
\right)
\end{equation}
where
\begin{eqnarray}
q_{-1}=-\left(\frac{1}{m^2}+\sum_{i=0}^{N}
\frac{2^{1-\delta_{i,0}}}{d_i^2}\right)^{-1/2},\quad
  q_{j}=-\left(\frac{1}{m^2}+\sum_{i=0}^{N}
\frac{2^{1-\delta_{i,0}} d_i^2}{(d_i^2-m_{V\,(j)}^2)^2}\right)^{-1/2}.
\end{eqnarray}
 Let us remark that the rotations to obtain
the gauge field mass eigenstates, $P_V$,
are different from those of the would-be GB, $Q_V$.
Consequently there could be a modification to the
ET that relates amplitudes of longitudinal
mass eigenstate gauge bosons,
${\bf \hat V^\mu}=P_V^t \, {\bf  V^\mu}$,
with those containing would be GB,
${\bf \hat G^V}=Q_V^t\,{\bf G^V}$
in the $R_\xi$ gauges.
As it is well known, the ET follows from the gauge-fixing Lagrangian 
\cite{ET,Lee:1977eg},
which, in terms of mass eigenstates, is now written as
\begin{eqnarray}
{\cal L}_{GF}(x)&=&-\f{1}{2\xi}\Big\{2\left\vert
\partial_\mu {\bf \hat W^{+ \mu}}-\xi P_W^t M^W_\xi Q_W{\bf \hat G^+}
\right\vert^2
+\left\vert
\partial_\mu {\bf \hat Z^\mu}-\xi P_Z^t M^Z_\xi Q_Z{\bf \hat G^Z}
\right\vert^2
\nonumber\\
&& 
\qquad+\left\vert \partial_\mu {\bf A^{\mu }}-\xi
M^A_\xi\,\,{\bf G^A}\right\vert^2 \Big\}
\end{eqnarray}
In this way it may seem that the $n$ mode of the gauge field
eigenstates $\hat V_{(n)}^\mu$ could mix  with all the $\hat
G^V_{(m)}$. Amazingly, the $q_i$ are related to 
the $u_i$ (also for finite $N$):
\be
q_i=
\f {d_0^2-m_{V(i)}^2}{\sqrt{2}{ m_{V(i)}}}u_i,
 \label{norm}
 \ee
which allows us to write:
\be P^t_V M_{V\xi} Q_V=\left(
\begin{array}{c|ccc}
0&m_{V(0)}&0& \ldots\\
0&0&m_{V(1)}&\ldots\\
\vdots&\vdots&\vdots&\ddots
\end{array}
\right)
\ee
and therefore there is no mixing
between the $N+1$ gauge bosons $\hat V_{(n)}^\mu$ and
the $N+2$ GB $\hat G^V_{(m)}$ unless
$n=m$. In other words, the longitudinal components of
the $\hat V_{(n)}^\mu$ will ``eat'' only the corresponding
$\hat G^V_{(n)}$, which is an eigenstate of the gauge-dependent
GB mass matrix. In particular, the $\hat G^V_{(-1)}$
are not GB combinations ``eaten'' by the longitudinal gauge bosons,
but remain in the physical spectrum as the physical charged and
neutral pseudoscalars. We can thus write simply:
\begin{eqnarray}
{\cal L}_{GF}(x)=   -\f{1}{\xi}\sum_{n=0}^{\infty}&&
\left\{ \f{1}{2}\left[ \partial_\mu A^{\mu }_{(n)}-\xi
{\f{n}{R}}\,\,A_{(n)}^5\right]^2\right.+\left\vert
\partial_\mu \hat W^{+\,\mu }_{(n)}-
\xi m_{W(n)}\hat G^+_{(n)}\right\vert^2\nonumber\\
&& \left. + \f{1}{2}\left[ \partial_\mu \hat Z^{\mu }_{(n)}-\xi
m_{Z(n)}\hat G^Z_{(n)}
\right]^2\right\}.
\end{eqnarray}
Once identified the $\hat G^V_{(n)}$ fields that couple diagonally with 
the derivatives of the gauge boson mass eigenstates,
the ET proof proceeds as usual \cite{ET,Lee:1977eg},
simply by substituting $V_L\rightarrow\hat V_L$ and
the would-be GB by $\hat G^V$. Therefore we arrive at
\begin{equation}
  T(\hat V^{\mu }_{L\,(m)},\hat V^{\mu }_{L\,(n)},\ldots)
\simeq  C^{(m)} C^{(n)}... T(\hat G^V_{(m)},\hat G^V_{(n)},\ldots) +O(M_k/E)
\end{equation}
$M_k$ being the biggest one of the 
$m_{V(m)},m_{V(n)}...$ masses, and the $C^ {(i)}=1+O(g)$
account for renormalization corrections (see the last three references 
in \cite{ET}).

\section{The 5D SM with one Higgs on the brane }

Let us study first the simple case of a single Higgs on the brane,
which is obtained from the general case by taking the
$c_\beta,\lambda^{(5)}_i\rightarrow0$, $i=1,3,4,5$,
$\mu_1\rightarrow\infty$ limit. As an application of the ET we
will illustrate how to calculate the
$\hat W^+_{L(m)}\hat W^-_{L(n)}\rightarrow \hat W^+_{L(p)}\hat W^-_{L(q)}$ 
amplitudes,
with $m,n,p,q \ge0$, which are thus related to $T_{mnpq}=\hat
G^+_{(m)}\hat G^-_{(n)}\rightarrow\hat  G^+_{(p)}\hat G^-_{(q)}$.
Among other things, these amplitudes are interesting to obtain
bounds on the Higgs masses from tree level unitarity. Similarly to
what it is done to obtain the unitarity limits in the SM, we are
only interested in the lowest order calculation in the gauge couplings 
 $g$ and $g'$.

We have decoupled the $\omega$ fields, so that
$G^\pm_{(0)}\rightarrow0$  and
${\bf G^\pm}=(\pi^\pm,0,W_{5\,(1)}^\pm,W_{5\,(2)}^\pm...)^t$.
Moreover, since 
${\bf \hat G^\pm}=Q_W^t {\bf G^\pm}$, and in this case  $d_0\rightarrow0$,
$q_{(-1)}/d_0\rightarrow-1$ so that
\begin{equation}
 \hat G^\pm_{(-1)}\rightarrow0, \quad \hat G^\pm_{(i)}=q_{i}\left(\f{1}{m_W}\pi^\pm+\sqrt{2}\sum_{n=1}^{N}
\f{n/R}{(n/R)^2-m_{W(i)}^2}W_{5\,(n)}^\pm\right),\quad i\ge 0
\end{equation}
Furthermore the scalar
potential now only depends on $\Phi_2$.
After integration on the 5th dimension, the relevant
coupling terms for the amplitude above are the usual
$\lambda_2 (\pi^+\pi^-)^2+2\lambda_2 v_2\pi^+\pi^-h_2$.
Note that in this case it is enough to look for
couplings of the $\pi$ fields since
there is no coupling of $W_{5\,(n)}^\pm$ gauge field
components to the $\Phi_2$ scalar sector in
$(D_\mu \Phi_2)^\dagger D^\mu \Phi_2$, but only a mixing
from the gauge fixing. Thus by substituting
$\pi^\pm=\sum_{i=0}^{N} (q_{i}/m_W) \hat G^\pm_{(i)}$
in the coupling terms, we find, for 
$\sqrt{s}\gg m_{W(m)},m_{W(n)},m_{W(p)},m_{W(q)}$
\begin{equation}
  T_{mnpq}=-i\sqrt{2}\,G_F \,m_{h_2}^2
\frac{q_m}{m_W}\frac{q_n}{m_W}\frac{q_p}{m_W}\frac{q_q}{m_W}
\left( \frac{s}{s-m_{h_2}^2}+ \frac{t}{t-m_{h_2}^2}\right).
\end{equation}

In particular,  for the scattering of longitudinal zero modes,
we find
the very same SM amplitude \cite{Lee:1977eg}, but corrected by a factor
\begin{eqnarray}
  \left(\f{q_0}{m_W}\right)^4=
\left(\f{2}{(1+\pi^2R^2m_W^2)+m_{W\,(0)}^2/m_W^2}\right)^2
\simeq 1-\f{2}{3}m_W^2\pi^2R^2+O(m_W^4 R^4),
\label{eq:0000}
\end{eqnarray}
where we have used eqs.~(\ref{norm}) and (\ref{ujdef}).
In  the last step we have also used the small $R$ approximation
$m_{W(0)}^2\simeq m_W^2(1-\pi^2R^2m_W^2/3)$
obtained by expanding the $c_\beta\rightarrow0$ limit
of the eigenvalue equation in eq.~(\ref{eigenvalueeq}).
As long as $R>(3\,$TeV$)^{-1}$,  
the corrections are rather small: $O(m_W^2 R^2)\simeq 10^{-3}$. 

Nevertheless, we next show that the modification
from the four gauge boson amplitudes can be even smaller.
As a matter of fact, the complete study of the unitarity bounds involves
amplitudes also with the Higgs or the
$W^\pm$ and $Z$  gauge bosons.
In particular, one is interested in the largest eigenvalue of the
matrix made of all these amplitudes.
The complete analysis of the unitarity bounds lies therefore 
beyond our applications of the ET.
However the ET will  allow us to  calculate the block
of $T_{mnpq}$ amplitudes  
in the $s,t\rightarrow \infty$ limit
\begin{eqnarray}
\left(
    \begin{array}{ccccc}
T_{0000}&T_{0010}&T_{0001}&T_{0011}&\ldots\\
T_{1000}&T_{1010}&T_{1001}&T_{1011}&\ldots\\
T_{0100}&T_{0110}&T_{0101}&T_{0111}&\ldots\\
T_{1100}&T_{1110}&T_{1101}&T_{1111}&\ldots\\
\vdots&\vdots&\vdots&\vdots&\ddots
    \end{array}
\right)
\stackrel{s\rightarrow\infty} {\longrightarrow}
-i\f{4 m^2_{h_2} G_F}{\sqrt{2}m_W^4}
\left(
    \begin{array}{ccccc}
q_0^4&q_0^3q_1&q_0^3q_1&q_0^2q_1^2&\ldots\\
q_0^3q_1&q_0^2q_1^2&q_0^2q_1^2&q_0q_1^3&\ldots\\
q_0^3q_1&q_0^2q_1^2&q_0^2q_1^2&q_0q_1^3&\ldots\\
q_0^2q_1^2&q_0q_1^3&q_0q_1^3&q_1^4&\ldots\\
\vdots&\vdots&\vdots&\vdots&\ddots
    \end{array}
\right)
\label{Tmatrix}
\end{eqnarray}
It can be shown that the largest eigenvalue is 
$-i4 m^2_{h_2} G_F(q_0^2+q_1^2+q_2^2+...)^2/(\sqrt{2}m_W^4)$ and
therefore the SM value is modified by a factor
\begin{eqnarray}
\label{largeeigenbulk}
\left(\sum_{k=0}^\infty \f{q^2_k}{m_W^2}\right)^2&=&
\left(\f{q_0}{m_W}\right)^4
+2  \left(\f{q_0}{m_W}\right)^2
\frac{1}{m_W^2}\sum_{k=1}^\infty q^2_k+...  \\
&&\simeq
1-\f{2}{3}m_W^2\pi^2R^2+2\sum_{k=1}^\infty
\frac{2}{1+\pi^2R^2m_W^2+ k^2/(m_W^2 R^2)}+ O(m_W^4R^4) \nonumber\\
&&\simeq
1-\f{2}{3}m_W^2\pi^2R^2+4 m_W^2 R^2 \sum_{k=1}^\infty
\frac{1}{k^2}+ O(m_W^4R^4)\simeq 1+ O(m_W^4R^4)
\nonumber
\end{eqnarray}
where we have approximated $m_{W\,(k)}\simeq k/R$ for $k\ge1$
and small $R$.

We stress that,  as long as $s\gg m_{W(n)}^2, m_{Z(n)}^2$, 
the same matrix pattern of eq.~(\ref{Tmatrix}) occurs for any
other four-gauge boson amplitude matrices, and therefore,
at least in the gauge sector, the same strong cancellations up
to $O(m_W^4 R^4)$ will occur.

\section{The 5D SM with one Higgs in the bulk }

Let us then study the other limiting case
when there is only one Higgs in the bulk,
which  is obtained from the general case by taking the
$s_\beta,\lambda_2, \lambda^{(5)}_i\rightarrow0$, $i=3,4,5$,
$\mu_2\rightarrow\infty$ limit. Note that $m=0$ and therefore all the
${\cal M}_V^2$ and ${\cal M}_{V\xi}^2$ matrices are already diagonal.
Physically, this means that there is no mixing
between any KK mode of different KK level.
 Thus, everything is simpler
since $\hat V= V$ and $\hat G^V= G^V$, with $c_\beta=1$ in eq.~(\ref{Gs}).

As before, we will
calculate the
$\hat W^+_{L(m)}\hat W^-_{L(n)}\rightarrow \hat W^+_{L(p)}\hat W^-_{L(q)}$
amplitudes, with $m,n,p,q \ge0,$ which, at high energies,
are related through the ET with
$T_{mnpq}=G^+_{(m)}G^-_{(n)}\rightarrow G^+_{(p)}G^-_{(q)}$.

Once more we are only interested in the lowest order calculation
in $g$ and $g'$ and thus we do not need the $\omega\omega V$ couplings.
Therefore, the only relevant interactions come from the scalar potential
and are given by
\begin{eqnarray}
&&\lambda_1\Big\{
(\omega^-_{(0)}\omega^+_{(0)})^2+
\omega^-_{(0)}\omega^-_{(0)}\omega^+_{(n)}\omega^+_{(n)}+
\omega^+_{(0)}\omega^+_{(0)}\omega^-_{(n)}\omega^-_{(n)}+
4 \omega^-_{(0)}\omega^+_{(0)}\omega^-_{(n)}\omega^+_{(n)}+\nonumber\\
&&\left.
+\sqrt{2} (\omega^-_{(0)}\omega^+_{(n)}\omega^-_{(m)}\omega^+_{(p)}+h.c.) \Delta_3(m,n,p)
+\f{1}{2}\omega^-_{(n)}\omega^+_{(m)}\omega^-_{(p)}\omega^+_{(q)} \Delta_4(m,n,p,q)
\right.\nonumber\\
&&\left.
+2 v_1 (\omega^-_{(0)}\omega^+_{(0)}h_{1\,(0)}+\omega^-_{(n)}\omega^+_{(0)}h_{1\,(n)}
+\omega^-_{(0)}\omega^+_{(n)}h_{1\,(n)}+\omega^-_{(n)}\omega^+_{(n)}h_{1\,(0)})\right.
\nonumber\\
&&
+\f{v_1}{\sqrt{2}} \omega^+_{(m)}\omega^-_{(n)}h_{1\,(p)}\Delta_3(m,n,p)
\Big\}
  \label{eq:potentialbulk}
\end{eqnarray}
where
we have used the usual convention of a summation over any repeated index,
with $n,m,p,q\ge1$.
In addition,
\begin{eqnarray}
&& \Delta_3(m,n,p)= \delta_{m+n}^{p}+\delta_{m+p}^{n}+\delta_{m}^{p+n}\\
&&  \Delta_4(m,n,p,q)= \delta_{m+n+p}^{q}+\delta_{m+n+q}^{p}+
\delta_{m+p+q}^{n}+
\delta_{m+n}^{p+q}
+\delta_{m+q}^{n+p}
+\delta_{m+p}^{n+q} +\delta_{m}^{n+p+q}. \label{eq:deltas}\nonumber
\end{eqnarray}
In principle, for our calculation we should recast the above expressions in terms
of the mass eigenstates, which in this case are the $G$ 
and $a$ fields in eqs.(\ref{Gs}) and (\ref{as}).
However, since there is no $\omega^+\omega^-\omega^3$ coupling, 
and that of $\omega^+\omega^- Z$
is of higher order, there is no $G^+G^- a^Z$ coupling at leading order.
Therefore, it is enough for our purposes to substitute in eq.~(\ref{eq:potentialbulk}):
$\omega^\pm_{(0)}\rightarrow -G^\pm_{(0)}$,
$\omega^\pm_{(n)}\rightarrow s^{_W}_{n} \, G^\pm_{(n)}$,
and read the $G^\pm$ coupling directly. 
Note that $T_{0000}$ is exactly the same
as that of the SM, but 
all the other $T_{mnpq}$ amplitudes are $O(m_W^2 R^2)$, since they contain at least
two $s^{_W}_n$. 
However,
it is possible to show that the corrections from the longitudinal
gauge sector to the tree level unitarity bounds on the Higgs mass
are indeed $O(m_W^4R^4)$. Indeed,
the dominant terms at $s,t\rightarrow\infty$ are given by the quartic couplings,
because the other diagrams are suppressed by $h_1$ propagators,
thus:
{
\begin{equation}
\left(
    \begin{array}{ccccc}
T_{0000}&T_{0010}&T_{0001}&T_{0011}&...\\
T_{1000}&T_{1010}&T_{1001}&T_{1011}&...\\
T_{0100}&T_{0110}&T_{0101}&T_{0111}&...\\
T_{1100}&T_{1110}&T_{1101}&T_{1111}&...\\
\vdots&\vdots&\vdots&\vdots&\!\!\ddots
    \end{array}
\right)
\stackrel{s\rightarrow\infty} {\longrightarrow}
\f{-i4 m^2_{h_1} G_F}{\sqrt{2}}
\left(
\setlength{\arraycolsep}{.5mm}
    \begin{array}{ccccc}
1 &0&0&(s^{_W}_1)^2&...\\
0&(s^{_W }_1)^2&(s^{_W }_1)^2&0&...\\
0&(s^{_W }_1)^2&(s^{_W }_1)^ 2&0&...\\
(s^{_W}_1)^2&0&0&3(s^{_W }_1)^4/2&...\\
\vdots&\vdots&\vdots&\vdots&\!\!\ddots
    \end{array}
\right),
\label{Tpmpmbulk}
\end{equation}
}
which, for small $R$,
 has a characteristic polynomial (in the $N\times N$ case),
$(1-\lambda)\,(\lambda^{(N-1)}+O(m_W^2R^2)\lambda^{(N-2)}+\ldots)+O(m_W^4R^4)\lambda^{N-2}=0$,
and hence the largest eigenvalue is 
$-i4 m^2_{h_1} G_F/(\sqrt{2})(1+O(m_W^4R^4))$,
the others are $O(m_W^2R^2)$
or zero. Therefore, only considering states of 
the Kaluza-Klein gauge sector, the corrections 
to the tree level SM unitarity bounds
are $O(m_W^4R^4)$. That is less than $10^{-6}$
for models where $R>(3 \hbox{TeV})^{-1}$
\cite{Masip:1999mk,Muck:2001yv}.
Even in the case when all fields live in the bulk
\cite{Barbieri:2000vh}, when $R$ can be as small as $300\,$GeV,
the corrections from these states could not be larger than $1\%$.

As a matter of fact, the full unitarity analysis should be carried out
also with the $Z_{(0)}$, $h_{1(0)}$ and their 
KK excitations, as well as the $a^V_{(n)}$ fields.
However, all other matrix amplitudes for two-gauge-boson scattering have
the same structure so that we find
again a very tiny correction to largest eigenvalue from the
pure Kaluza-Klein gauge sector blocks.
The amplitudes involving Higgs or  $a^V_{(n)}$ fields
are not calculated with the ET and lie beyond
the scope of this work.

\section{One Higgs in the bulk and one on the brane}

Let us now study the complete potential
in eq.~(\ref{scalarpotential}), with the scalar field $\Phi_1$ in
the bulk and $\Phi_2$ on the brane, using the full
formalism and notations given in section \ref{section2}. Once
more, as an application of the ET,  
we will study the
 $\hat W^+_{L(m)}\hat W^-_{L(n)}\rightarrow \hat W^+_{L(p)}\hat W^-_{L(q)}$
amplitude, which, again, is related to 
$T_{mnpq}=\hat G^+_{(m)}\hat G^-_{(n)}\rightarrow\hat  G^+_{(p)}\hat G^-_{(q)}$. 

As in the previous cases, the dominant unitarity violation
in the $s,t\rightarrow\infty$  limit 
is given by the quartic $\hat G^\pm_{(n)}$ 
couplings from eq.~(\ref{scalarpotential}). They are obtained by rewriting
$\pi^\pm,\omega^\pm_{(n)}$ in terms of $G^\pm_{(n)}$, inverting
eqs.(\ref{Gs}) and (\ref{as}), and then using ${\bf G^\pm}=Q_W {\bf\hat  G^\pm}$.
This amounts to the following substitutions:
\begin{eqnarray}
&&  \pi^\pm \rightarrow \frac{q_0}{m}\,\hat  G^\pm_{(0)}
+\frac{q_1}{m}\,\hat  G^\pm_{(1)}+\ldots
\nonumber \\
&&\omega^\pm_{(0)}\rightarrow -\frac{d_0\, q_0}{d_0^2-m^2_{W(0)}}
\,\hat  G^\pm_{(0)}
-\frac{d_0\, q_1}{d_0^2-m^2_{W(1)}}\,\hat  G^\pm_{(1)}-\ldots
\nonumber \\
&&\omega^\pm_{(n)}\rightarrow \sqrt{2} \,s^{_W}_n \Big[
\frac{d_n\, q_0}{d_n^2-m^2_{W(0)}}\,\hat  G^\pm_{(0)}
+\frac{d_n\, q_1}{d_n^2-m^2_{W(1)}}\,\hat  G^\pm_{(1)}+\ldots
\end{eqnarray}
In this way we have reexpressed the potential in terms of
the would-be GB:
$\hat  G^\pm_{(0)},\hat  G^\pm_{(1)}\ldots$. Note that
we are not interested in quartic couplings with $\hat  G^\pm_{(-1)}$
because it is not a would-be GB and is not ``eaten'' by any
longitudinal gauge boson. After some tedious but straightforward calculations,
we find, up to $O(m_W^2R^2)$:
\begin{eqnarray}
\left(
    \begin{array}{ccccc}
T_{0000}&T_{0010}&T_{0001}&T_{0011}&\ldots\\
T_{1000}&T_{1010}&T_{1001}&T_{1011}&\ldots\\
T_{0100}&T_{0110}&T_{0101}&T_{0111}&\ldots\\
T_{1100}&T_{1110}&T_{1101}&T_{1111}&\ldots\\
\vdots&\vdots&\vdots&\vdots&\ddots
    \end{array}
\right)
\stackrel{s\rightarrow\infty} {\longrightarrow}
-i
\left(
    \begin{array}{ccccc}
Aq_0^4      &Bq_0^3q_1  &B q_0^3q_1  &C q_0^2q_1^2&\ldots\\
Bq_0^3 q_1  &Cq_0^2q_1^2&C q_0^2q_1^2&0&\ldots\\
Bq_0^3 q_1  &Cq_0^2q_1^2&C q_0^2q_1^2&0&\ldots\\
Cq_0^2q_1^2 &0          &0           &0&\ldots\\
\vdots&\vdots&\vdots&\vdots&\ddots
    \end{array}
\right)
\label{Tmatrixgeneral}
\end{eqnarray}
where, 
\begin{eqnarray}
  \label{eq:ABC}
  A=\frac{4\,\lambda_1\,d_0^ 4}{[d_0^2-m^2_{W(0)}]^4}+\frac{4\,\lambda_2}{m^2},
\quad B=\frac{4\,\lambda_2}{m^2},\quad C=\frac{4\,\lambda_2}{m^2}
+\frac{2\, \lambda_1 c_\beta^4}{m^4 s_\beta^8}+O(m_W^2 R^2)
\end{eqnarray}
The largest eigenvalue (compare with eq.~(\ref{largeeigenbulk})) is now given by :
\begin{eqnarray}
   && \label{eq:largegeneral}
A q_0^4+\frac{2 B^2}{A}q_0^2\sum_{i=1}^\infty q_i^2+O(m_W^4 R^4)\simeq
\\ && \qquad\qquad\frac{4\,G_F}{\sqrt{2}}\Big[m_{h_1}^2c_\beta^2+m_{h_2}^2s_\beta^2\Big]
\Big\{1+\frac{2\pi^2 s_\beta^4c_\beta^4}{3}(m_W R)^2
\frac{[m_{h_1}^2-m_{h_2}^2]^2}{[m_{h_1}^2c_\beta^2+m_{h_2}^2s_\beta^2]^2}
+O(m_W^4 R^4)\Big\}
\nonumber
\end{eqnarray}
Hence, in the general case we find that the strong cancellation
of the simple cases studied before, which are recovered in the
$s_\beta\rightarrow$ and $c_\beta\rightarrow0$ limits, does not occur,
and, unless $m_{h_1}=m_{h_2}$, there is an $O(m_W^2R^2)$ 
modification to the SM result from the pure gauge sector.

\section{Conclusions}

In this work we present a generalization of the Equivalence Theorem 
between longitudinal gauge bosons and Goldstone bosons to the case
when there is one extra dimension  and the Standard Model gauge symmetry is spontaneously
broken by a Higgs field in the bulk and another one on the brane.
The main difficulty is the identification of the would be Goldstone bosons,
which are a mixture of the familiar Goldstone bosons with their own Kaluza
Klein excitations and those of the fifth component of the gauge bosons.

The Equivalence Theorem turns out to be
a powerful tool to obtain simple expressions involving
longitudinal gauge bosons as we have illustrated by
calculating their scattering amplitudes in several cases.
The ET has allowed us to show that the modifications
from pure longitudinal gauge boson scattering to the tree level unitarity
bounds of the SM are generically small 
and in the one Higgs limiting cases can suffer from even stronger cancellations.

Our results open up the possibility to tackle the full matrix 
needed for the complete unitarity
violation study, including also amplitudes involving Higgs fields
\cite{noi}.

\end{document}